\documentclass[a4paper]{spie}  %>>> use for US letter paper
\usepackage{amsmath,amsfonts,amssymb}
\usepackage{graphicx}
\usepackage[colorlinks=true, allcolors=blue]{hyperref}

\title{R\&D activity in INAF on light instrumentation for continuous optical turbulence monitoring}
\author[a,c]{Camilo Weinberger}
\author[a,c]{Elena Masciadri}
\author[a,c]{Luca Carbonaro}
\author[a,c]{Marlene de Sepibus}
\author[b,c]{Fernando Pedichini}
\author[a,c]{Ciro Del Vecchio}
\author[d]{Christian Veillet}
\author[d]{Sam Ragland}

\affil[a]{INAF-OAA, Largo Enrico Fermi 5, 50125 Firenze}
\affil[b]{INAF-OAR, Via Frascati 33, Monte Porzio Catone, Roma}
\affil[c]{ADONI: Adaptive Optics National Laboratory in Italy}
\affil[d]{LBTO 933 N. Cherry Drive, Tucson, AZ, USA, 85721}

\authorinfo{C.W.: E-mail: camilo.weinbergercerro@inaf.it\\  E.M.: E-mail: elena.masciadri@inaf,it}

\begin{document} 
\maketitle

\begin{abstract}
Measurements of optical turbulence are essential in both ground-based astronomy and free-space optical communication systems contexts for multiple reasons including supporting and optimizing Adaptive Optics. 
Similarly, forecasting these parameters is highly valuable for planning observational campaigns; however, it requires at least two years of continuous data to generate a reliable training dataset.  
The most commonly used instruments to monitor continuously the optical turbulence include DIMM, MASS, SLODAR, RINGSS sensors.
The Arcetri team is leading two operational forecast systems for two different Observatories: ALTA above LBT, and FATE above VLT. Both systems are in continuous evolution and obviously forecast performances depend also on available measurements.  Our long-term goal is to develop our own instruments to obtain continuous measurements directly from local observations.

In this work, we present an automated system designed for continuous seeing measurements using multiple sensors. This portable station integrates a Cyclope, a RINGSS unit, a scintillometer which operate together to provide comprehensive night-time atmospheric information. We will compare the performance of low-cost sensors, such as the Cyclope and the scintillometer, with that of the RINGSS turbulence profiler, which has been fully installed and programmed by our team in collaboration of LBT team. A dedicated insturment for night and day-time measurements is planned for near future.

This system incorporates a fully automatic on-site system that monitors operational status, delivers real-time information on sensor conditions and measurements, and manages synchronization with the data storage unit. The system’s real-time acquisition capability enables continuous turbulence analysis, supports R\&D activities, facilitates local data generation, and allows on-site testing of forecasting methodologies.
\end{abstract}

%\keywords{R\&D, Instrumentation, Seeing, RINGSS, turbulence measurement, Turbulence profiler, Automatization.}

\section{INTRODUCTION}
\label{sec:intro}  % \label{} allows reference to this section

Optical turbulence (OT) measurements are a fundamental element for the optimization of the OT forecasts in astronomy \cite{masciadri2020,masciadri2023} as well as in free-space optical communication (FSOC), and forecasts are crucial for the planning of astronomical observations as well as of networks managing optical ground stations (OGS) in the FSOC projects. OT measurements are also useful to support adaptive optics (AO) in both contexts and they can even be retrieved from AO systems \cite{quatresooz2026}. In perspective it will be more and more interesting to enlarge the sky temporal coverage of measurements to the day time and there are already studies that went in that direction \cite{griffiths2023,song2020,chabe2020}. 

Recent work of our team\cite{weinberger2026_1,weinberger2026_2} underscores how turbulence forecasting done with neural networks strongly depends on the access of data-set of measurements extended on long period of time preferably as continuous as possible. For this reason we decided to start a R\&D activity aiming to develop portable and automatized instruments that can be used as individual instruments or as a multi-sensor OT measurements station. This system integrates low-cost sensors a Cyclope plus a CSSMS (Continuous Solar Seeing Measurement System) both of them from Alcor System\footnote{\href{https://www.alcor-system.com/}{https://www.alcor-system.com/}} alongside a RINGSS \cite{tokovinin2021,bustos2025} turbulence profiler. Forthcoming instruments dedicated to a comprehensive day- and night-time atmospheric characterization. This R\&D activity has as objectives:
\begin{itemize}
    \item (1) Replicate instruments proposed in the literature in order to collect large data-set of observations to be used for machine learning approach for the OT forecast (e.g. RINGSS or Cyclope). The goal is to be able to automatized and replicate the instrument to be placed in different locations.
    \item (2) Investigate the level of performances/validity of simple and low-cost techniques such as Cyclope (for night time) and the scintillometer (CSSMS for day time) frequently used by the community (particularly within the context of free-space optical communications (FSOC)). 
    \item (3) The activity described at points 1 and 2 will allow us to characterise the techniques we have mentioned to support our collaborators at LBTO that recently implemented a Cyclope and a RINGSS (located @PEPSI telescope) to enrich the typology of astroclimatic parameters retrievable \cite{veillet2026}. This instrumentation will be fundamental to optimize the operations and the new generation AO system at LBTO \cite{ragland2026_1, ragland2026_2}. Due to logistical constraints LBTO is forced, indeed, to locate these instruments in specific locations without the freedom to compare their measurements with other references and without the possibility to quantify potential biases that we can do with our R\&D. 
    \item (4) Development of new instruments. At present we are focusing our attention on the development on an instrument able to cover the OT estimate during night and day time (24H). We submitted a request for funds.
\end{itemize}

\begin{figure}[h]
    \centering
    \includegraphics[width=0.75\linewidth]{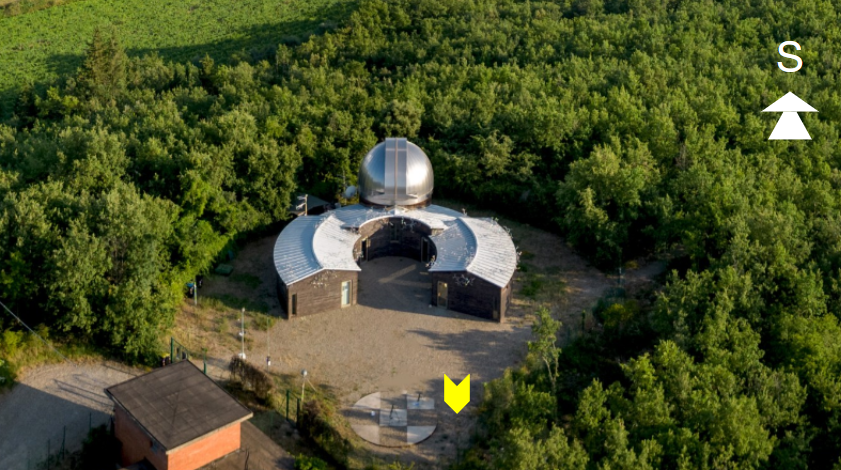}
    \caption{Osservatorio del Chianti. The instruments are supposed to be run continuously for testing and validation at the Osservatorio Polifunzionale del Chianti (OPC, Figure 2), located 36 km from Florence. The yellow arrow indicates the position where to locate instrumentation}
    \label{fig:place}
\end{figure}

\section{Hardware}
    This station works as modular system connecting each instrument with the others in case one wishes a simultaneous estimation of the optical turbulence from all of them and also to protect the instruments from the adversal condition during the measurement. The whole system is connected to a main control system  using a RaspBerry Pi5 that manages the information from each module to the server and the data storage. Currently we have 4 principal modules for the R\&D system: Cyclope, a CSSMS (Continuous Solar Seeing measurement system) that is a scontillometer, a RINGSS and the Weather station WXT530 developed by Vaisala. The selection and the  development of these instrument is for R\&D therefore each module has to be able also as an individual autonomous module.

    This work contemplates a foldable astro-meteorological station aimed to provide continuously information of the seeing, $C_N^2$ and derived integrated astroclimatic parameters ($\tau_{0}$ and $\theta_{0}$), pressure, temperature, wind speed, wind direction and rain information that runs 24/7 to create a full information on site. 
    The weather station provides information to control automatically the floodgate that protect instruments from adverse ambient conditions.
    
    %\color{blue}(attenzione Camilo, non vogliamo caratterizzare l'OPC ma gli strumenti. Ho modificato la frase leggermente) \color{black}
    The most suitable site where to locate our instrumentation for the R\&D is at present the Osservatorio Polifunzionale del Chianti (OPC), located at $36$km away from Florence, Italy (Figure \ref{fig:place}). This site provides a full sky view over $20^o$ of elevation, optimal for operative goals.
    %The weather recordings will be supervised by WXT530 Weather Station from Vaisala and connected to the system by a RaspBerry Pi 5. 
    %For the turbulence sensors we acquired three sensors dedicated to various task for R\&D aims: Cyclope, CSSMS and RINGSS.
    \color{black}

\section{Instruments for optical turbulence measurements}

\subsection{Cyclope and Scintillometer}

\begin{figure}[h]
    \centering
    \includegraphics[width=0.75\linewidth]{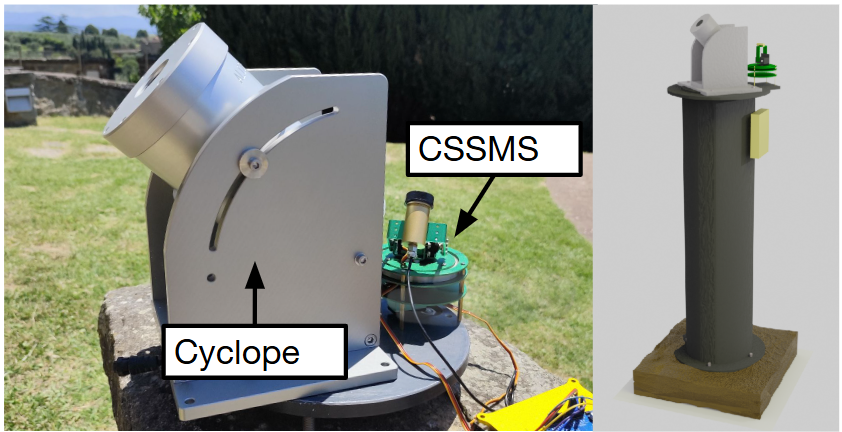}
    \caption{Mounting for Cyclope (pointing to the north)  and the CSSMS (pointing to the south)}
    \label{fig:cyclope}
\end{figure}

    One of the module incorporates two low-cost sensors developed by Alcor Systems (Figure \ref{fig:cyclope}). Cyclope seeing monitor, is a streamlined version of a DIMM that estimates the integration of the OT from the wavefront angle of arrival using a single aperture targeted at Polaris. Although this architecture trades off some accuracy, it has not yet been fully validated against other standard sensors. Cyclope is aligned with the Earth's rotational axis to track the Polar. For daytime monitoring, the module utilizes the CSSMS, a solar scintillometer that measures variations in solar irradiance and incorporates a dedicated servo system for active sun tracking. No validations exist at present for this instrument too. Figure \ref{fig:cyclope}-left show this module.

    This first module corresponds to a module running continuously 24H recording seeing, integrated by two sensors strategically located and coordinated to obtain the seeing measurement during the night (Cyclope) and during the Day (scintillometer).
     Due to the fact that Cyclope does not provide a differential measurement as DIMM (where we subtract angle of arrivals coming from two holes) the Cyclope is sensitive to vibrations of the instrument. In order to eliminate such vibrations it is necessary to fix Cyclope on top of a dedicated structure blocked by a concrete base (Figure \ref{fig:cyclope}-right). Finite element analysis has been performed to define the gemetrical characteristics of the tube. This sensor has a FoV of $3.6^o\times 2.5^o$ and track Polaris by software, for the same reason does not require a mount for tracking.
    The day-time sensor correspond to the Continuous Solar Seeing Measurement System (CSSMS) from Alcor--systems, this sensor estimates the equivalent value of the seeing from the scintillation of the sun.

\begin{figure}[h]
    \centering
    \includegraphics[width=.6\linewidth]{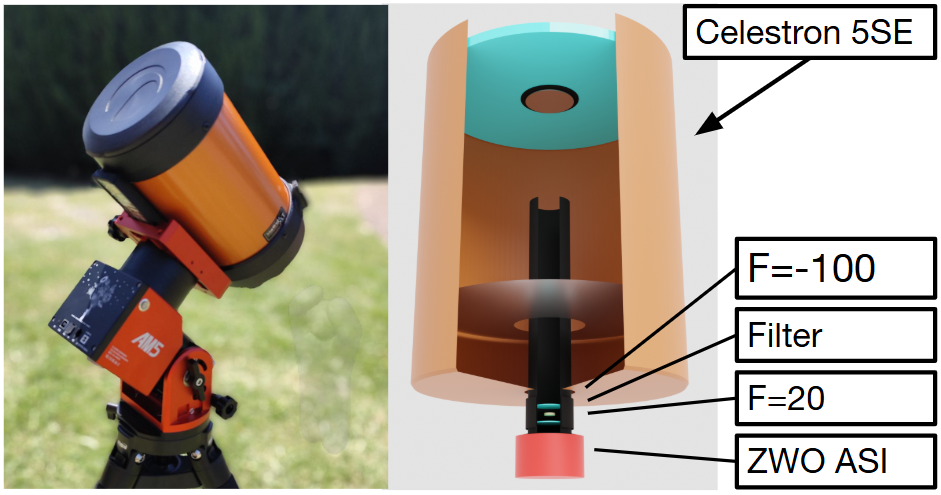}
    \caption{Left: RINGSS turbulence profiler. Right: principal pieces forming the RINGSS.
}
    \label{fig:ringss}
\end{figure}

\subsection{RINGSS}
\begin{figure}[h]
    \centering
    \includegraphics[width=0.65\linewidth]{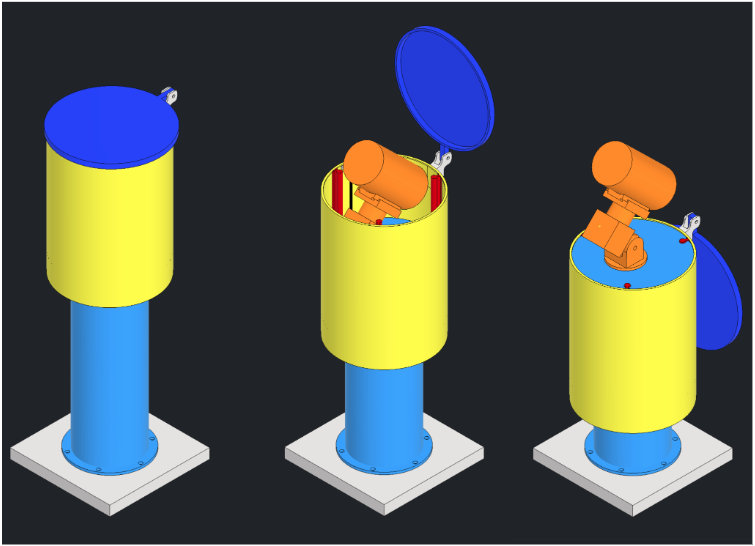}
    \caption{Schematic view of the dome conceived for the RINGSS. In the automatized version of the RINGSS the instruments has to be able to be stopped and covered in case the ambient conditions do not allow the instrument to run properly. The operative conditions are: wind speed below 20 m/s, not raining (rain sensor), sun position 18 degrees below the horizon.}
    \label{fig:dome}
\end{figure}

The second module incorporates the RINGSS sensor (Figure \ref{fig:ringss}), which provides optical turbulence profiles at low resolution on 20 km using ring-shaped intensity maps captured on a defocused sensor. RINGSS is a low vertical resolution profiler providing 8 layers reaching the top of the atmosphere at which we are interested on ($\sim$ 20 km). Differently from the MASS, RINGSS is not blind to the turbulence close to the ground. To safeguard the instrument against environmental conditions, it features a specialized, compact dome that operates autonomously based on weather station data (Figure \ref{fig:dome}). This dome is developed by our team. Automated star tracking and pointing are driven by an integrated ZWO AM5 mount.

    \begin{table}[ht]
    \caption{RINGSS principal components.} 
    \label{tab:Multimedia-Specifications}
    \begin{center}       
    \begin{tabular}{|l|l|}
    \hline
    \rule[-1ex]{0pt}{3.5ex}  Component & Description  \\
    \hline
    \rule[-1ex]{0pt}{3.5ex}  Celestron Nexstar 5SE & Telescope $D=12.5$cm\\
    \hline 
    \rule[-1ex]{0pt}{3.5ex}  ZWO ASI 290MM & High \textit{snr} camera\\
    \hline 
    \rule[-1ex]{0pt}{3.5ex}  AM5N Harmonic Equatorial & Electronical monture\\
    \hline 
    \rule[-1ex]{0pt}{3.5ex}  SVPRO 5-50	 & Pointing camera\\
    \hline 
    \end{tabular}
    \end{center}
    \end{table}

\section{Software}

    The automation architecture is developed on Python 3.11, integrating serial communication and housekeeping routines to monitor sensor and storage server status. While Cyclope and CSSMS rely on Alcor Systems' native software, our software synchronizes system updates during sunrise and sunset intervals to avoid interfering with active data acquisition. For tracking, in the case of RINGSS, an algorithm was developed using the Skyfield library and the Hipparcos catalog to identify stellar positions. This allows for comprehensive sky coverage analysis, identifying optimal guide stars based on magnitude and position constraints. 

\begin{figure}[h]
    \centering
    \includegraphics[width=0.6\linewidth]{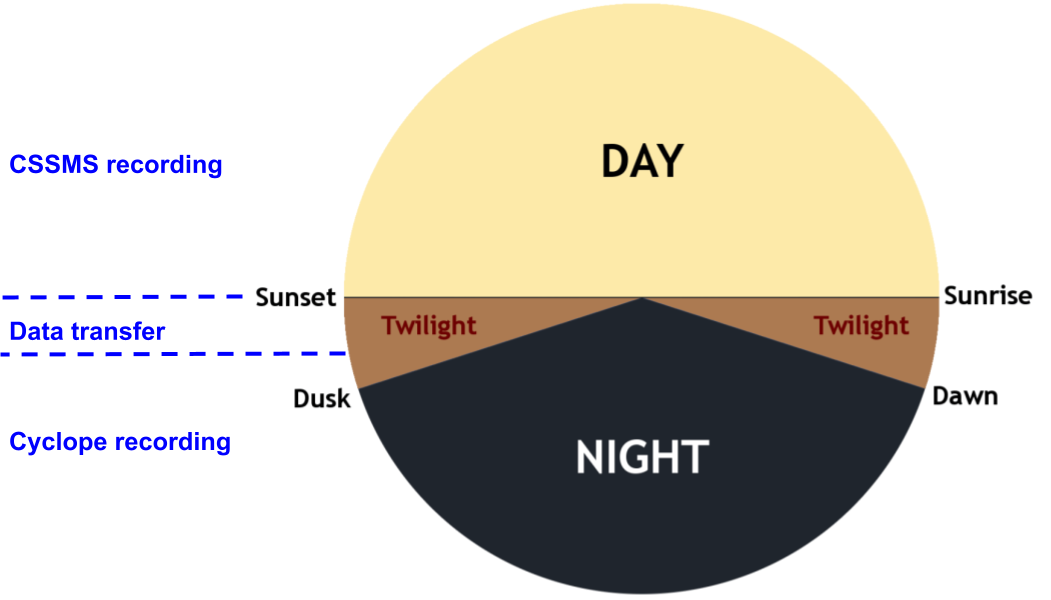}
    \caption{This figure displays the operation of the system depending of the part of the day. There are two small intervals of time close to the beginning of the night and the day in which there are no measurements. The instruments do not run when the sun is below 5 degrees with respect to the horizon. During these two intervals data are transferred to the Arcetri's stored repositories.}
    \label{fig:record_transfer}
\end{figure}

    The software works with parallel modules that can communicate one with the other and that constantly check the status of the codes of the individual instruments and the status of the files-outputs produced by the individual instruments. The most relevant parts of the software are:
    \begin{itemize}
        \item The main code: it provides the measurements that are transmitted in real time to the website allowing a remote control. The main code controls the data-flux from the sensors to the data--store server in Arcetri.
        This code continuously checks the status of the weather-station and other sensors on real time.
        \item Weather station control: this program works as a flag system for RINGSS, because the high wind-speed and rain could generate problems to the sensors of instruments.
        \item Control of RINGSS dome and mounting: this code check the flags from the weather condition and moves the RINGSS position to the target or to the fold position depending of the sun position and the flag.   
        \item RINGSS estimator: This code (based on the original Tokovinin's code), controls the camera, estimates the astroclimatic parameters and generates the files containing the measurements that are transmitted on local storage systems each day.
        \item Cyclope and CSSMS software (Windows): both softwares are developed by Alcor-systems and work in similar way. Cyclope covers the night time and the CSSMS covers the day time. The acquisition starts when the Sun position is at -5$^{\circ}$ with respect to the horizon therefore there are two small intervals of time  that are close the beginning of the day and the beginning of the night in which there are no measurements as indicated in Fig. \ref{fig:record_transfer}. The system is done in a way that in these intervals of time the observations done in the night or in the day are transmitted to the Arcetr's stored repositories. 
    \end{itemize}

    Having different instruments running in parallel we can easily perform comparisons among observations (if wished) and or tests to set-up the instruments or even validate them. The main goal of this action is that to develop a set of instruments that can be eventually replicated to be placed where we wish to feed forecasts systems such as machine learning and neural networks or AR.

\section{Sky coverage and automatization}
    
     We calculated first the sky coverage for the two sites (Florence and Mt.Graham). A premise has to be done. All estimates of the seeing done along whatever line of sight have to be corrected by the air mass. This correction is based on the fact that the seeing along a line of sight having a zenithal angle $\zeta$ ($\varepsilon$) is related to the seeing estimated along the zenith ($\varepsilon_{0}$) by the formula: 
     \begin{equation}
        \varepsilon_{0}=\varepsilon \times { \sec (\zeta )}^{3/5},
        \label{eq1}
    \end{equation}
     The correction by the air mass is necessary because we need to normalize all measurements with respect to the zenith. The sky coverage is the percentage of stars that we can observe respecting observations constraints that are in part imposed by the observed stars (for RINGSS the constraint is m $\le$ 2) and by geometrical constraints on the zenithal angle $\zeta$ (typical an external constraint). In Florence there are no geometrical constraints from external factors therefore the final constraints are m $\le$ 2 and  $\zeta$  $\le$ 35$^{\circ}$. We have indeed interest in observing stars that are as close as possible to the zenith. In this way it is more difficult not to respect the approximation of weak perturbations.
     For the Mt.Graham case (therefore the site of the LBT), we have however an external constraint due to the fact that RINGSS is placed close to the LBT dome and the limitations in viewing are $30^o \le \zeta \le 60^o$ for the zenithal angle and for the azimuth $-85^o \le \theta \le 120^o$ (where $\theta$=0 is the South direction). We calculated that we have a sky coverage equal to 100\% in both sites for all the periods of the year. That means that we should not have problems in finding stars to be observed with RINGSS in whatever night of the year. 
  
     Once the sky coverage is guarantee the most crucial issue is the implementation of the automatization of the instrument. The heart of the automatization is based on the identification and tracking of stars by respecting few conditions that maximize the efficiency of the observations: \\ \\
     (1) selection of the most brillant star therefore miniminzing m if more than one option is available \\
     (2)  maximizing the time of observation of the identified star in order to avoid to change too frequently the star. The interrumption of the observations leads to  too frequent gaps in the observations \\ 
     (3)  minimizing the zenithal angle  \\
     
\begin{figure}
    \centering
    \includegraphics[width=0.75\linewidth]{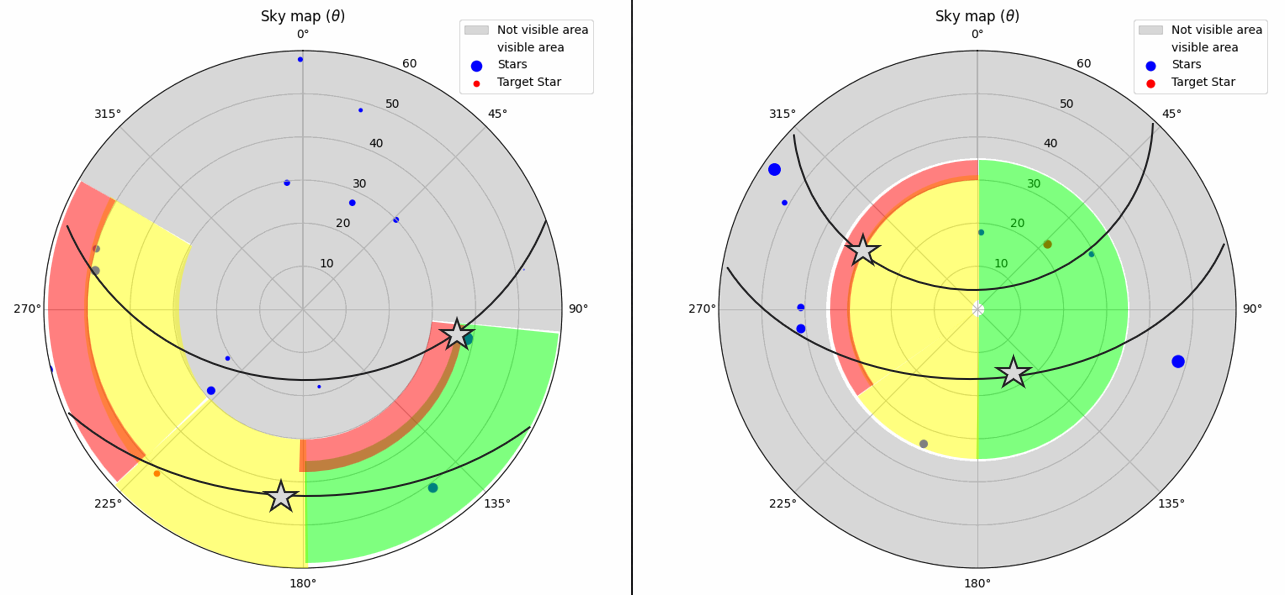}
    \caption{This figure represents the star selection criteria based on the constrains from each location and the stars magnitude. Mt. Graham (left) and Paranal (right). The priority is based on color: on green (high priority) those stars that assert more than 2 hours of continuous measurement, yellow: at least 1 hours of continuous measurement, red: less than 1 hour.
}
    \label{fig:prior}
\end{figure}

     Once established these conditions the code of the automatization is developped taking care to define order of priority in the selection of the star. An example of the logic followed by the automatic procedure is represented in Fig. \ref{fig:prior} showing the sky polar maps of the observations above Florence and Mt.Graham. The table of color indicates the order of priority where the highest priority in green and lowest priority in orange. The code will focus on the stars coming from the East in first place, as it shown in the Figure \ref{fig:prior}.
     
     Another aspect to take into account in the automatic process is the moon position because the moon generates an additional background noise near to stars. For the same reason, we included a filter that eliminates all the stars within $10^o$ away from the moon.
     
%\begin{equation}
%    cos(\Delta) \le sin(\zeta_1)\cdot sin(\zeta_1) + cos(\zeta_2)\cdot cos(\zeta_1)\cdot sin(\theta_2-\theta_1)    
%\end{equation}

\section{Conclusion and forthcoming work}

Our team recently decided to start a R\&D activity focused on portable and light instruments conceived for monitoring the optical turbulence during the night and/or day. The main goals are described in the Introduction: advancing instrumentation R\&D, enabling rigorous cross-instrument validation, training high-performance neural networks for turbulence forecasting in astronomy and FSOC and supporting our collaborators from LBTO who also purchased a Cyclope and a RINGSS with multi-goals that go from providing to ALTA a wide typology of observations for the optimization of the science operation and to optimize the AO systems at LBTO.

We started this activity focusing our attention on three instruments: RINGSS, Cyclope and CSSMS (a scintillometer) and we looked for funding to develop a new instrument covering the 24H (day and night).

We can conclude that we identified a site, we identified the infrastructure on which to implement each individual instrument, we conceived a dedicated dome specifically drawn for RINGSS, we identified an optimal support for Cyclope that is able to kill mechanical vibrations, we purchased the dedicated sensors that are able to control the ambient conditions and protect the individual instruments when the instruments are run in an automatic configuration, we calculated the sky coverage above Florence and Mt.Graham and results tell us that we have a 100$\%$ sky coverage above both sites and we developed a software (written in Python 3.11) aiming to make the system automatic. Such a system ensures optimal target selection, guaranteeing uninterrupted data acquisition to mitigate seasonal gaps.  

The forthcoming work is that to start to run operatively these instruments (Cyclope and CSSMS are at a more advanced level of progress with respect to RINGSS) and to investigate more in detail the RINGSS estimates: seeing and vertical stratification.

\acknowledgments % equivalent to \section*{ACKNOWLEDGMENTS}       
We acknowledge the Large Binocular Telescope Observatory (LBTO) for their constant support as well as the support from the LBTO Director Joe Shields.

% References
%\bibliographystyle{mnras}
%\bibliography{report} % bibliography data in report.bib
\bibliographystyle{spiebib} % makes bibtex use spiebib.bst

\end{document}